\documentclass[12pt]{article}
\usepackage{graphicx}%
\usepackage{graphics}
\usepackage{bm}

\usepackage{enumerate}
\usepackage{float}
\usepackage{dsfont}

\usepackage{amsmath}
\usepackage{amsfonts}
\usepackage{amssymb}
\usepackage{natbib}

\usepackage{color}
\usepackage[displaymath]{lineno}
\newcommand*\patchAmsMathEnvironmentForLineno[1]{%
  \expandafter\let\csname old#1\expandafter\endcsname\csname #1\endcsname
  \expandafter\let\csname oldend#1\expandafter\endcsname\csname end#1\endcsname
  \renewenvironment{#1}%
     {\linenomath\csname old#1\endcsname}%
     {\csname oldend#1\endcsname\endlinenomath}}%
\newcommand*\patchBothAmsMathEnvironmentsForLineno[1]{%
  \patchAmsMathEnvironmentForLineno{#1}%
  \patchAmsMathEnvironmentForLineno{#1*}}%
\AtBeginDocument{%
\patchBothAmsMathEnvironmentsForLineno{equation}%
\patchBothAmsMathEnvironmentsForLineno{align}%
\patchBothAmsMathEnvironmentsForLineno{flalign}%
\patchBothAmsMathEnvironmentsForLineno{alignat}%
\patchBothAmsMathEnvironmentsForLineno{gather}%
\patchBothAmsMathEnvironmentsForLineno{multline}%
}
\usepackage{setspace}
\newcommand{\E}{\mathbb{E}}
\renewcommand{\P}{\mathbb{P}}
\newcommand{\Var}{\mathbb{V}ar}

\newcommand{\Comb}[2]{\left(\begin{array}{c}#1 \cr #2\end{array}\right) }
\begin{document}
\title{Extracting abundance indices from longline surveys: a method to account for hook competition and unbaited hooks}
\author{Marie-Pierre Etienne\footnote{AgroParisTech/INRA UMR 518, 16 rue Claude Bernard, F-75231 Paris, France}, Shannon G. Obradovich\footnote{Fisheries Centre, Aquatic Ecosystem Research Laboratory, 2202 Main Mall, The University of British Columbia,  Vancouver,  BC V6T 1Z4, Canada},\\
 K. Lynne Yamanaka\footnote{Pacific Biological Station, 3190 Hammond Bay Road, Nanaimo,BC V9T 6N7, Canada} and  Murdoch K. McAllister\footnotemark[2]}
\maketitle

\begin{abstract}
To estimate fish population trends and abundance, fisheries scientists commonly apply dynamic population models fitted to relative abundance indices.  Populations are often monitored using longline fishing gear and the most commonly used relative abundance index in this case is the catch per unit effort (CPUE), defined  as the number of fish of the targeted species caught per hook and minute of soak time. Longline CPUE can be affected by interspecific competition and the retrieval of unbaited or empty hooks, and this can lead to biases in the apparent abundance trends.  Interspecific competition has been previously studied but the return of empty hooks is ignored in all current treatments of longline CPUE.  
This work proposes and compares different stochastic models to define indices to address both issues simultaneously. Maximum likelihood estimators and their asymptotic covariance matrices are obtained. Simulating different joint scenarios for interspecific competition and the empty hooks, we show that CPUE behaves badly in every scenario. Information about the source of the empty hooks is required to select the appropriate identifiability constraint and therefore derive the appropriate abundance index. The above methods are applied to build relative indices from 2003 to 2009 for quillback rockfish ({\it Sebastes maliger}) in British Columbia from longline survey data.  Due to variation in the incidence of non-target species, the index trend obtained is moderately sensitive to the choice of the estimator. The proposed methodology permits the building of reliable abundance indices for all populations monitored using longline fishing gear.
\end{abstract}

\section{Introduction}

Many fish stock assessments derive information about stock trends from analyses of catch and effort records obtained from a longline fishing event \citep{Clark+06,Maunder+10}. In this paper, we define the event of setting a longline with baited hooks for some fixed amount of time, hauling the longline and recording the location and the species caught per hook as a ``longline experiment''.   
The classical relative abundance index, Catch Per Unit Effort (CPUE), for populations monitored using longline gear, is defined as the average number of individuals of targeted species caught per hook and minute of soak time.  This commonly used CPUE index of abundance ignores the variability introduced by the competition for baited hooks within and between species. \cite{Somerton95} propose two different indices, based on instantaneous rate of catch from longline survey records  which takes interspecific competition into account. \cite{Haimovici+07} claim that CPUE and the instantaneous rate give the same results. \cite{Ward+04} shows with another model that soaktime and without taking competition into account that soaktime is important. The authors also proved that other factors may influence the indices but the effects of those factors are not addressed in this paper.  Another important source of variability in the abundance indices which has received relatively little attention is the presence of empty hooks, i.e.,  hooks returning without bait or fish. 

This paper generalizes the use of instantaneous rates of catch from longline surveys as relative abundance indices (as proposed by \cite{Somerton95} and  \cite{Rothschild67}) and evaluates alternative approaches to dealing with  empty hooks in the formulation of longline survey stock trend indices.

The paper is divided into three main sections after the introduction. Section \ref{sec:review} reviews existing methods for deriving relative abundance indices from longline catch and effort records. Section \ref{sec:EmptyHooks} presents our generalization of the previous methods to account for empty hooks. Some simulation studies are conducted to compare the indices under different levels of interspecific competition and sources for empty hooks.  Section \ref{sec:Results} gives the results of the simulation studies and illustrates the behaviour of the indices for monitoring the abundance of quillback rockfish ({\it Sebastes maliger}). The full technical details are outlined in the appendix.

\section{Review of methods to derive abundance indices from longline records}
\label{sec:review}
\subsection{ Catch Per Unit Effort (CPUE)}
In a longline survey CPUE is defined as the ratio between the number of fish  of the target species caught, $N_T$, and  the number of hooks,  $N$, times the soak time, $S$:
\begin{equation*}CPUE= \frac{\# Target}{\# hooks \times Soak Time} =\frac{N_T}{N \, S} .\end{equation*}
This index ignores the effects of competition and  gear saturation.

\subsection{The simple exponential model} 
\citet{Somerton95} have proposed two  alternative  approaches to deal with the issues of hook competition and gear saturation. Only one is useful in our context because dealing with gear saturation requires observed capture times which are not often recorded as in our case study.  The number of available baits on the longline is considered to decrease at an exponential rate measuring the overall pressure on the hooks.  This overall pressure may be split into a sum of relative abundance indices per species. Using this approach the  $k^{th}$ catch $C_{r,k}$ for a soak time $S$ and for species $r=1,\ldots R$ is given by:
\begin{equation}
\label{eq:SEmDef}
C_{r,k}=\frac{\lambda_r}{\lambda} N(1-e^{-\lambda S} )+\varepsilon_{r,k}, \quad\mbox{with}\quad \varepsilon_{r,k}\overset{i.i.d}{\sim}\mathcal{N}(0,\sigma^2),
\end{equation}

where $N$  stands for the initial number of hooks, $\lambda$ is the overall pressure on hooks, and $\lambda=\sum_{r}\lambda_r$ is the sum of the specific abundance index per species, and $k$ is the number of the set. The parameters $(\lambda_1,\ldots,\lambda_R)$ define a relative abundance index for each $r$ species\footnote{ Mostly one particular species is the target species, let $r=1$ while the other species, $r>1$ are non-target species. In this context it is easier to consider only $\lambda_1=\lambda_T$ the relative abundance for the target species and $\lambda_2=\lambda_{NT}$ the relative abundance index which summarizes all the other species.}.

A few of the drawbacks of this  Simple Exponential Model (SEM) are the assumptions of normality and homoscedasticity of the error terms. It is intuitive that the variability should be higher for species with higher relative abundance, i.e., the variance of the error term should not be constant but should depend in some way on $\lambda_t$. Furthermore $C_{r,k}$ is a discrete number (number of fish caught), potentially small and the normal assumption is not accurate in this case.

\subsection{Multinomial Exponential Model}
This section reviews the underlying ideas  of \citet{Somerton95} and proposes an alternative  model which more closely  mimics the behavior of the fish. This model has been originally proposed by  \cite{Rothschild67} and describes how the catch of a target species could be reduced by the catch of other species.

Let us define $T_T$ as the time it takes to catch an individual from the target species on one particular hook. $T_T$ is assumed to follow an exponential distribution of rate $\lambda_T$, i.e
\begin{equation*}\P \left(  T_T \geq u \right) = e^{-\lambda_T u},\end{equation*}
where $u$ is some fixed amount of time from when the baited hook was placed in the species' habitat. $T_{NT}$ is an exponential random variable with parameter $\lambda_{NT}$ and models the time it takes to catch an individual from any of the non-target species. We can define $T=\min\left\lbrace T_{T}, T_{NT}\right\rbrace$ as the time it takes to catch an individual of any species. Thanks to the property of the exponential distribution, $T$ is exponentially distributed with rate $\lambda=\lambda_T + \lambda_{NT}$. This property justifies the decomposition of the overall relative abundance as a sum of specific abundance given by \citet{Somerton95}. \\

After the soak period $S$, for one hook, there are only three possible outcomes:
\begin{itemize}
\item ${\left \lbrace I=0 \right \rbrace} = \left \lbrace \mbox{ The hook is still baited.}\right\rbrace$. It means that the time for a capture is greater than the soak time. This event occurs with probability
  \begin{equation*}\P(I=0)=\P(T>S)={e^{-\lambda\, S}}\end{equation*}
\item  ${\left \lbrace I\ne0 \right \rbrace} = \left \lbrace \mbox{ The hook is no longer baited.}\right\rbrace$.
  \begin{equation*}\P(I\ne 0)=\P(T<S)={1-e^{-\lambda\, S}}.\end{equation*}
Given the hook is no longer baited, there are two possible outcomes: the catch is either from the target species, i.e.
$\left \lbrace I=T \right \rbrace $,which occurs with probability
    \begin{equation*}
      \P(I=T)  ={\P(  T_T<T_{NT} \vert T<S )}{\P(  T<S )} = {\frac{\lambda_T}{\lambda}}{ (1-e^{-\lambda\, S})},
    \end{equation*}
    or the catch is from a non-target species, corresponding to the event $\left \lbrace I=NT \right \rbrace$ which  occurs with probability:
    \begin{equation*}
      \P(I=NT) ={\P( T_{NT}<T_T \vert T<S)}{ \P( T<S  )} ={\frac{\lambda_{NT}}{\lambda}} {(1-e^{-\lambda\, S})}.
    \end{equation*}
  \end{itemize}

Assuming that all the hooks on a longline behave independently,  the likelihood is given by
\begin{equation}
\label{eq:lik1}
L(\lambda_T, \lambda_{NT}) =
\left (
  \begin{array}{c} N
    \\ N_B
  \end{array}
\right)  \left (
\begin{array}{c}
  N_{T}\\
  N_{T}+N_{NT}
  \end{array}
\right) (e^{-\lambda\, S})^{N_B}  (1-e^{-\lambda\, S})^{N_{T}+N_{NT}} \left(\frac{\lambda_T}{\lambda}\right)^{N_{T}} \left(\frac{\lambda_{NT}}{\lambda}\right)^{N_{NT}},
\end{equation}
where
\begin{itemize}
\item $N$ is the number of hooks on the longline,
\item $N_B$ is the number of baited hooks at the end of the soak time,
\item $N_{T}$ is the number of individuals of the target species caught,
\item $N_{NT}$ is the number of individuals of the non-target species caught.
\end{itemize}
The combinatorial terms arise since all the hooks are considered independent and the order of the catch on the longline has no importance. This model was originally proposed by \citet{Rothschild67} although presented here with a slightly different approach. This model is called Multinomial Exponential Model (MEM) since the vector $(N_B, N_T, N_{NT})$ follows a multinomial distribution whose vector of probability depends on an exponential term.
\bigskip

If $\lambda_{NT}$ is larger than $\lambda_T$, it corresponds to a high level of competition: for a given relative abundance of the target species $\lambda_T$, the catch decreases as $\lambda_{NT}$, the non-target species relative  abundance, increases.

\subsection{Links between the indices}
\subsubsection{ Links between MEM and SEM}
The expected number of fish caught of the target species $N_T$ is the same under the MEM and  SEM assumptions and is given by:
\begin{equation*}\E \left ( N_{T} \right) = N \frac{\lambda_T}{\lambda} \left(1- e^{-\lambda S}\right).\end{equation*}

Moreover, the models share the same parameters, $\lambda_T$ and $\lambda_{NT}$. The main difference is the error term. In the SEM, the error term is normally distributed with a variance given by:
\begin{equation*}\Var_{SEM}(N_T)=\Var(N_{NT})=\sigma^2,\end{equation*}
while in the MEM the total number of fish caught has a multinomial  distribution and the variance is given by:

\begin{equation*}\left \lbrace \begin{array}{l}
\Var_{MEM}(N_T)=N \frac{\lambda_T}{\lambda} (1-e^{-\lambda S})  \left  ( 1 - \frac{\lambda_T}{\lambda} (1-e^{-\lambda S}) \right) \\
 \Var_{MEM}(N_{NT})=N \frac{\lambda_{NT}}{\lambda} (1-e^{-\lambda S})  \left  ( 1 - \frac{\lambda_{NT}}{\lambda} (1-e^{-\lambda S}) \right) .
\end{array}\right.
\end{equation*}
Furthermore $N_T$ and $N_{NT}$ are assumed to be independent in the SEM and not in the MEM since they are drawn from the same multinomial probability distribution.
\bigskip

\subsubsection{Links between CPUE and MEM}
 Under the MEM assumption, the expected CPUE of the target species is given by:
\begin{equation*}
\E \left (CPUE \right)= \frac{\lambda_T}{\lambda\, S} \left(1-e^{-\lambda\, S}\right)=\lambda_T + o(\lambda)
\end{equation*}
If the overall relative density index $\lambda$ is small enough, meaning that there is little competition and the target species is not very abundant, CPUE and the MEM index give the same results. This theoretical result is consistent with the expected behavior. Furthermore the limit when the soaktime goes to 0 is equal to $\lambda_T$, which means that the relative abundance index is the equvalent of instantaneous CPUE. 

\section{Dealing with empty hooks}
\label{sec:EmptyHooks}
In longline experiments, it is common for some hooks to return empty; the hook is no longer baited, but there is no fish on it.  There could be several explanations for  these empty hooks such as mechanical removal of bait during gear setting/retrieval, consumption of the bait by invertebrates or  fish without being hooked or removal of the hooked fish by predators.

\medskip
In this paper,  we consider the hypothesis that  empty hooks arise only from the escape of fish. Therefore, the question about empty hooks is reduced to ``How should  the empty hooks be allocated to the different species?'' These empty hooks provide information that we could use to improve the quality of our abundance indices. This section  describes modifications of  the MEM to incorporate this information and details some statistical properties of the indices built using these versions of MEM. It  also describes different ways to include the empty hook information into the SEM.

\subsection{Full version of the Multinomial Exponential Model}
 We propose a modified version of the Multinomial Exponential Model  to account for empty hooks. As opposed to the previous version of the MEM, here each fish caught has a probability of escaping equal to $p_T$, for target species, and $p_{NT}$, for non-target species. \\

\begin{figure}[h!]
\includegraphics[width=8cm]{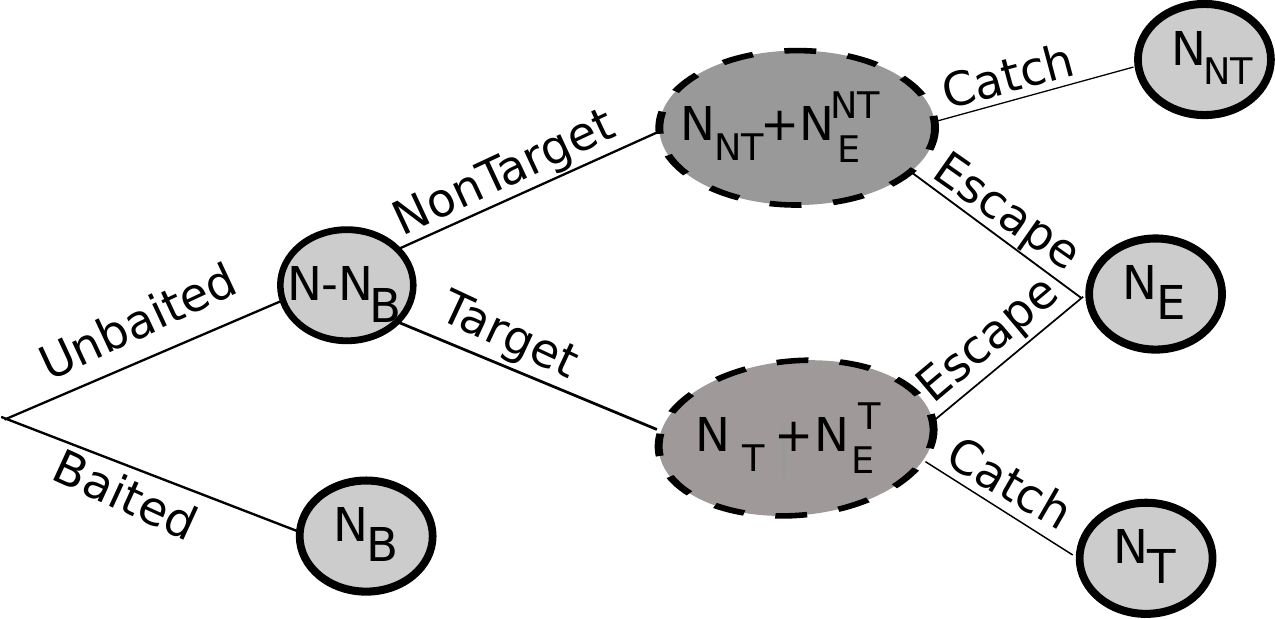}
\caption{Conditional description of the model. The observed quantities are solid lines, the hidden quantities are dashed lines.}
\label{Fig:1-Outcome}%
\end{figure}

 We use three additional variables to fully specify the model: $N_E$ is the number of observed empty hooks; $N_E^{(T)}$ (respectively $N_E^{(NT)}$) stands for the number of target species (respectively  non-target species) individuals  which have escaped and these two random variables are not observed.   Assuming, as for the simple version of MEM,  that all hooks are independent, we are able to conditionally describe the outcomes (Figure \ref{Fig:1-Outcome}).
\begin{itemize}
 \item  The number $N_B$ of baited hooks retrieved at the end of the soak time is the realisation of a binomial random variable with probability of success $e^{-\lambda S}$.  
$$ N_{B} \sim \mathcal{B}\left(N, e^{-\lambda S}\right).$$
\item Among the $N-N_B$ empty hooks, the total number of individuals from target species caught is $N_T + N_E^{(T)}$ and is also binomially distributed:
$$N_T+  N_E^{(T)}\vert N_B \sim \mathcal{B}\left(N-N_B, \frac{\lambda_1}{\lambda}\right).$$
\item Given $N_T+ N_E^{(T)} $, the total number of individuals from target species caught and landed on board  is $N_T$ and is also binomially distributed:
$$N_T\vert N_T +  N_E^{(T)} \sim \mathcal{B}\left(N_T +N_E^{(T)} , (1-p_T) \right).$$
\item Given $N_{NT}+ N_E^{(NT)} $, the total number of individuals from non-target species caught and landed on board  is $N_{NT}$ and also has a binomial distribution:
$$N_{NT}\vert N_{NT} +  N_E^{(NT)} \sim \mathcal{B}\left(N_{NT} +N_E^{(NT)} , (1-p_{NT}) \right).$$
\end{itemize}

The full version of Multinomial Exponential Model is summed up through a probability tree in Figure \ref{Fig:1-Outcome}. $N_E^{(NT)}$ and $N_E^{(T)}$ are missing quantities but the sum $N_E$ of these two quantities is observed. Appendix \ref{an:Distribution} gives the main step to define the likelihood of this model:
\begin{align}
\label{eq:MEMlogLike}
 l(\lambda_T, \lambda_{NT}, p_T, p_{NT} ) = & \frac{N!}{N_B !\, N_T ! \, N_{NT}!\, N_E ! } \left(e^{-\lambda S}\right)^{N_B}  \left(1-e^{-\lambda S}\right)^{N-N_B} \cr
&\left( \frac{\lambda_T}{\lambda}(1-p_{T})\right)^{N_T}\left( \frac{\lambda_{NT}}{\lambda}(1-p_{NT})\right)^{N_{NT}}\left( \frac{\lambda_{T}p_T + \lambda_{NT}p_{NT}}{\lambda}\right)^{N_{E}}
\end{align}

The full version of Multinomial Exponential Model may be considered as a multinomial distribution:
\begin{align*}
& (N_B, N_T, N_{NT}, N_{E}) \sim \mathcal{M}\left (N, \boldsymbol{\alpha} \right)\\
\mbox{with }  &\boldsymbol{\alpha} = \left(e^{-\lambda S},\  (1-e^{-\lambda S})\frac{\lambda_T}{\lambda} (1-p_T), \  (1-e^{-\lambda S})\frac{\lambda_{NT}}{\lambda} (1-p_{NT}),\   (1-e^{-\lambda S})\frac{\lambda_T p_T + \lambda_{NT}p_{NT}}{\lambda} \right)
\end{align*}

In this full version the model is not identifiable since an equivalent version could be expressed with only three parameters $\lambda$, $\lambda_T (1-p_T)$ and $\lambda_{NT} (1-p_{NT})$ (see annex \ref{an:Identifiable} for more details). Some additional information is  required to estimate the parameters in this model.  It is possible to add some biological knowledge  on the probability of escape through prior distribution in a Bayesian framework but no information of this kind is available for our case study. This point is discussed in section \ref{sec:Discussion}. Considering a frequentist  or an objective Bayesian approach some particular solutions have to be chosen. The idea is to put a constraint of identifiability which links $p_{NT}$ and $p_T$. This constraint may be expressed as $P_T=\alpha P_{NT}$. Any choice of $\alpha$ according to biological consideration may be relevant but in this paper we focus on two reasonable choices:
\begin{enumerate}
\item MEM1: empty hooks come only from non-target species, so $p_{T}$ is assumed to be 0, this corresponds to $\alpha=0$. Most of the time the target species is less abundant than all the non-target species. Allocating the empty hooks to the non-target species will at worst lead to an underestimation of the target species. Furthermore baited hooks are designed to catch and retain the target species.

\item MEM2: another reasonable choice is to assume that the probability of escape is the same for target and non-target species, i.e $p_T=p_{NT}$ and $\alpha=1$.  An empty hook has had the bait stolen by a fish but no information about the species of this fish is available so that the empty hooks are allocated according to the relative densities of each group.
\end{enumerate}

\subsection{Maximum likelihood estimation of MEM}
As all the longline sets are supposed to be independent, the complete likelihood is simply the product of the likelihood for each experiment given by formula \ref{eq:MEMlogLike}. At this stage we have to consider two different situations: variable soak times or similar soak times.

If the soak times are different for all the longline sets, no analytical formulae for the estimators is available and  the estimation step has to be performed using a non linear  optimization algorithm.

Mostly, the longline experiments have been designed to share the same soak time to reduce the causes of variation in the experiment. In this case, an analytical formula can be derived for MEM1 and MEM2 because all the information can be summed up through the vector $(N_{B+}, N_{T+},N_{NT+}, N_{E+})$ which corresponds to $(\sum_{l=1}^L N_{B_l},\sum_{l=1}^L  N_{T_l},\sum_{l=1}^L N_{NT_l},\sum_{l=1}^L  N_{E_l})$, where $l$ is the number of the longline set.

Even if the design of the experiment prescribes a constant soak time for each set, the actual soak time can differ slightly due to weather conditions or practical reasons. If the difference is not important, it is judicious to consider a single soak time (the mean for instance) to avoid the need of a numerical optimization which can produce some instability in the estimation.

As detailed in appendix \ref{an:Estimations}, the maximum likelihood estimators are given by:

\begin{equation}
\begin{array}{cc}
MEM1 & MEM2\\
\left\lbrace
  \begin{array}{rl}
    \hat{\lambda}_{T}&=\frac{N_{T+}}{N_+-N_{B+}}  \frac{1}{S}\log {\left( \frac{N_+}{N_{B+}}\right)}\\
    \hat{\lambda}_{NT}&=\frac{N_{NT+}+N_{E+}}{N_+-N_{B+}} \frac{1}{S} \log {\left(\frac{N_+}{N_{B+}}\right)}\\
   \hat{p}_{NT}&=\frac{N_{E+}}{N_{E+}+N_{NT+}},\quad   \hat{p}_{T}=0
  \end{array}
\right. & \left\lbrace
  \begin{array}{rl}
    \hat{\lambda}_{T}&=\frac{N_{T+}}{N_{T+}+N_{NT+}} \frac{1}{S}\log {\left( \frac{N_+}{N_{B+}}\right)}\\
    \hat{\lambda}_{NT}&=\frac{N_{NT+}}{N_{T+} + N_{NT+}}  \frac{1}{S}\log {\left(\frac{N_+}{N_{B+}}\right)}\\
    \hat{p}_{T}&=\frac{N_{E+}}{N_{E+}+N_{T+}+N_{NT+}}= \hat{p}_{NT}
  \end{array}
\right. \\
\end{array}
\label{eq:MLE}
\end{equation}

Maximum likelihood estimators are asymptotically unbiased and the covariance matrix is the inverse of the Fisher Information Matrix  \citep{Severini00}. If the total number of hooks $N$ is large enough, the joint distributions of these estimators can be approximated by a Multivariate Normal distribution. Asymptotically the  covariance matrix for MEM1  is given by:
\begin{equation}
\label{eq:CovMEM1}
Cov_{MEM1} = \frac{\lambda_T\lambda_{NT}}{N (1-e^{-\lambda S})}\left (
  \begin{array}{ccc}
    \frac{(1-e^{-\lambda S})^2}{S^2 e^{-\lambda S}\lambda^2} \frac{\lambda_T}{\lambda_{NT}} +  1 &
     \frac{(1-e^{-\lambda S})^2}{S^2 e^{-\lambda S}\lambda^2} - 1 & 0 \\
   \frac{(1-e^{-\lambda S})^2}{S^2 e^{-\lambda S}\lambda^2}  - 1 &   \frac{1-e^{-\lambda S}}{S^2 e^{-\lambda S}\lambda^2} \frac{\lambda_{NT}}{\lambda_{T}} +  1&  0 \\
    0 & 0 & \frac{p_{NT} (1-p_{NT}) \lambda }{\lambda_T \lambda_{NT}^2 } \\
  \end{array}
\right),
\end{equation}

and for MEM2 by
\begin{equation}
\label{eq:CovMEM2}
Cov_{\small MEM_2} = \frac{\lambda_T \lambda_{NT}}{N (1-e^{-\lambda S})}\left (
  \begin{array}{ccc}
    \frac{ (1-e^{-\lambda S})^2}{S^2 e^{-\lambda S}\lambda^2} \frac{\lambda_T}{\lambda_{NT}} +  \frac{1}{1-p}&
   \frac{(1-e^{-\lambda S})^2}{S^2 e^{-\lambda S}\lambda^2}  -\frac{1}{1-p}  & 0 \\
     \frac{(1-e^{-\lambda S})^2}{S^2 e^{-\lambda S}\lambda^2} -\frac{1}{1-p}&    \frac{(1-e^{-\lambda S})^2}{S^2 e^{-\lambda S}\lambda^2} \frac{\lambda_{NT}}{\lambda_T} +  \frac{1}{1-p}  &  0 \\
    0 & 0 & \frac{p (1-p)}{\lambda_T \lambda_{NT}} \\
  \end{array}
\right).
\end{equation}

The result given by \cite{Rothschild67} concerning the asymptotic variance for the simple version of MEM with $p_T=p_{NT}=0$ should be the same replacing, $p$ or $p_{NT}$ with 0 in the above formulae. Nevertheless, the two formulas are not compatible even if the estimators are. We suspect a mistake in the formula proposed by \cite{Rothschild67}.

\subsubsection{Bayesian framework for MEM}
The multinomial exponential model could also be estimated in a Bayesian framework. The full specification of  the Bayesian version of the MEM requires the definition of some  prior distributions for the parameters $(\lambda_T,\lambda_{NT}, p_{T}, p_{NT})$. The relative abundance index is always less than 1. Therefore, in our study, the priors have been chosen as poorly informative and independent.
\begin{equation*}
\lambda_{T} \sim \beta(0.1, 0.1) \quad \lambda_{NT} \sim \beta(0.1,0.1).
\end{equation*}
If  there is no informative prior on the probability of escape, the model is still non-identifiable. In a Bayesian framework, an identifiable model can be diagnosed since the posterior distribution is the same as  the prior distribution. Nothing has been learned from the data.

To obtain useful results, informative prior distributions could be defined using biological knowledge or a field experiment. This aspect hasn't been investigated in this work. We use the specific forms of the model (MEM1 or MEM2) to remove the problem of identifiability.

The estimation procedure has been implemented  using the JAGS (Just Another Gibbs Sampling) software \citep{JAGS}. An example of JAGS code is  provided in section \ref{sec:jagscodeMEM1}  for MEM1 and \ref{sec:jagscodeMEM2} for MEM2 and all other codes are available on request.

\subsection{Estimation for other indices}

In this section, we will discuss the estimation steps of all the indices previously presented and highlights the key points of the procedure.
\subsubsection{Multiple CPUE}

In a given region, you could have several sets of longlines deployed, so the estimation step requires the fitting of  one index using several observations.

If only one set is deployed, the CPUE is obtained by the catch over the soak time multiplied by the number of hooks. A simple generalization of this index is detailed in Appendix \ref{sec:cpue}  and proposes to compute CPUE as
\begin{equation*}CPUE=\frac{\sum_{l=1}^L N_{T_l}}{\sum_{l}  S_l N_l},\end{equation*}
where $l=1,\ldots,L$ stands for the longline set. If the number of hooks or the soak times are different, the generalized CPUE computed is a weighted average of all individual CPUE. A simple average is not sufficient, since for example, the average will attribute the same weight to an experiment with 200 hooks and to another experiment with only 50 hooks.

\subsubsection{Simple Exponential model}
In the following, two versions of SEM are derived depending on how empty hooks are considered:
\begin{enumerate}
\item SEM1: empty hooks are assumed to arise only from the non-target species and $N_E$ and $N_{NT}$ are pooled together.
\item SEM2: empty hooks are considered as a "third" species, and an additional relative abundance index is defined $\lambda_E$.
\end{enumerate}

\cite{Hovgard00} proposed to estimate $\lambda$ using 
\begin{equation*}
\hat{\lambda}_{Hov} =\frac{ -\log{ (N_{B}/ N)}}{S}.
\end{equation*}
If all longline sets share the same soak time and the same initial number of hooks $N$, the MLEs for this model are almost the same as for the MEM with the exception of empty hooks.
\begin{align*}
SEM1 & 
\left\lbrace
  \begin{array}{rl}
    \hat{\lambda}_{T}&=\frac{N_{T+}}{N_+-N_{B+}}  \frac{1}{S}\log {\left( \frac{N_+}{N_{B+}}\right)}\\
    \hat{\lambda}_{NT}&=\frac{N_{NT+}+N_{E+}}{N_+-N_{B+}} \frac{1}{S} \log {\left(\frac{N_+}{N_{B+}}\right)}\\
   \hat{\sigma}^2 &=\frac{1}{2L} \sum_{l=1}^L \left ( N_{T_l} - N \frac{\hat{\lambda}_T}{\hat{\lambda}} (1-e^{-\hat{\lambda} S} )\right)^2 + \left ( N_{NT_l} +N_{E_l}- N \frac{\hat{\lambda}_{NT}}{\hat{\lambda}} (1-e^{-\hat{\lambda} S} )\right)^2
   \end{array}
\right. \\
SEM2 & \left\lbrace
  \begin{array}{rl}
    \hat{\lambda}_{T}&=\frac{N_{T+}}{N_+-N_{B+}}  \frac{1}{S}\log {\left( \frac{N_+}{N_{B+}}\right)}\\
    \hat{\lambda}_{NT}&=\frac{N_{NT+}}{N_+-N_{B+}} \frac{1}{S} \log {\left(\frac{N_+}{N_{B+}}\right)}\\
    \hat{\lambda}_{E}&=\frac{N_{E+}}{N_+-N_{B+}} \frac{1}{S} \log {\left(\frac{N_+}{N_{B+}}\right)}\\
   \hat{\sigma}^2 &=\frac{1}{3L} \sum_{l=1}^L \left \lbrace \left ( N_{T_l} - N \frac{\hat{\lambda}_T}{\hat{\lambda}} (1-e^{-\hat{\lambda} S} )\right)^2 + \left ( N_{NT_l} - N \frac{\hat{\lambda}_{NT}}{\hat{\lambda}} (1-e^{-\hat{\lambda} S} )\right)^2 + \right.\\
   &\hspace{1cm}\left.\left ( N_{E_l} - N \frac{\hat{\lambda}_{E}}{\hat{\lambda}} (1-e^{-\hat{\lambda} S} )\right)^2\right\rbrace
  \end{array}
\right. \\
\label{eq:SEMMLE}
\end{align*}

To estimate $\sigma$ at least two longline sets are required; otherwise the estimates correspond to a perfect match and there is no additional variability.\\
We are only interested in building abundance indices for the target species and since the estimations of $\hat{\lambda}_T$ in SEM1 and SEM2 are the same, in the following we call this index SEM.

When the soak times or the initial number of hooks are different, a numerical optimisation algorithm has to be used to define the MLEs for SEM1 and SEM2. This approach should be avoided if possible due to numerical instability. From a practical point of view the \verb+nlm+ function available in \verb+R+ software  \citep{Rcite} behaves badly. In this work we directly optimize the log likelihood function using function \verb+optim+.\\
The analytical formulas for $\lambda_T$ are exactly the same, for SEM and MEM1: if soak times and the initial number of hooks are the same for all sets, there is no difference between these two indices concerning the relative abundance of the target species, which is not true for the non-target species. But this analytical formula is only valid for SEM when soak times and the initial number of hooks are the same while MEM only requires the same soak times.

\subsection{Simulation studies}
Bias and variability of the estimators are evaluated in this section through simulation studies. Several plausible scenarios have been studied to give some robust results and advice about the behaviour of all the indices.

Specifying  data generators (Operating Models) is the first step of simulation studies. In our specific case, we have at first glance, different solutions. We can use the Simple Exponential Model but this model doesn't simulate any empty hooks, and it provides non-integer catch values because of the normal hypothesis. The two other solutions (ie MEM1 and MEM2) will be used to study the different scenarios.

This choice of Operating Model gives obviously an advantage to MEM1 if the data were simulated with MEM1 and to MEM2 in the other case.

For one set of fixed parameters (that is $\lambda_T$, $\lambda_{NT}$, $L$ the total number of sets, $S$ the soak time, $N$ the number of hooks on a longline) 5000 fake datasets are generated and the corresponding estimated values for $\lambda_T$ and $\lambda_{NT}$ computed. A relative bias and a coefficient of variation are derived from these simulations. To study the impact of the estimation via a numerical algorithm, we also compute the estimators by maximizing the log likelihood using a non-linear optimization algorithm.

The values $\lambda_T$ and $\lambda_{NT}$ need to be chosen to reflect a plausible situation. In this work, four values of each parameter have been used which are $10^{-5}, 5.10^{-5}, 10^{-4}, 5.10^{-4}$ for $\lambda_T$ and $5.10^{-4}, 10^{-3}, 5.10^{-3}, 10^	{-2}$ for $\lambda_{NT}$, and all of the sixteen combinations of these two parameters have been addressed. The values have been choosen according the observed relative abundance derived from the Rockfish survey described in section \ref{sec:presdata}.

Three different scenarios for empty hooks have been simulated:
\begin{enumerate}[Sc.1)]
\item There are no empty hooks. Each hook has caught a fish. This situation corresponds to $p_{NT}=p_{T}=0$. The operating model is MEM with $p_{NT}=p_T=0$.
\item The ability to escape is the same across species. The operating model is MEM2 with a probability of escape set to $p_{T}=p_{NT}=0.2$.
\item non-target species are better at escaping. The probability of escape is set to $p_{NT}=0.2$ for the non-target species and to $p_{T}=0.02$ for the target species. In this case the oeprating model is a MEM but it doesn't fill the assumption of MEM1 or MEM2.
\end{enumerate}

The results of this simulation study are presented in section \ref{sec:resSim}.
\subsection{Description of the B.C. inshore rockfish longline survey}
\label{sec:presdata}
Since 2003, Fisheries and Oceans Canada has conducted an annual research longline survey in the Strait of Georgia with the fisheries research vessel { \bfseries  {\em CCGS Neocaligus}}.  Different regions of the Strait are covered each year, resulting in each statistical area (PMFA) being surveyed every two to three years.  A 2 km by 2 km grid is overlaid on all inshore rockfish habitat up to 100 m in depth, as determined using Canadian Hydrographic Service charts.  These blocks are stratified by depth into shallow (41-70 m) and deep (71-100m) and $8\%$ of the blocks, in a given statistical area, are randomly selected for fishing each year (see Lochead and Yamanaka 2007 for further details).

The snap-type longline gear consists of 1800 ft of leaded groundline with 225 circle hooks (13/0) spaced 8 ft apart. Each hook is attached to the snap by a 1.2 ft perlon gangion, crimped at both ends, and with a swivel at the hook.  The hooks are baited with Argentinean squid.  Soak time for each set is two hours and measured as the time from deployment of the last anchor when setting the gear to the retrieval of the first anchor on board when hauling.

As the gear is retrieved, the condition of each hook is recorded as returning with bait, with catch, empty (i.e. without bait or catch on the hook), or unknown, if the hook does not return.  Catch is recorded to the species level for both fish and invertebrates.

In this paper, we will focus on quillback, which is one of the numerous species of Rockfish. 
\section{Results}
\label{sec:Results}

\subsection{Simulation studies}
\label{sec:resSim}
In the simulation studies, we focus on two measures of quality for the models. The bias has been chosen since the goal of relative abundance indices consists in  accurately reconstructing population trends. Unbiased abundance indices ensure a good trend reconstruction. The standard deviation of the indices is used as a second measure of quality to compare the models since a precise estimate is always welcomed.  

\subsubsection{Competition but no empty hooks}
\begin{center}
\begin{figure}[h!]
\includegraphics[width=11cm]{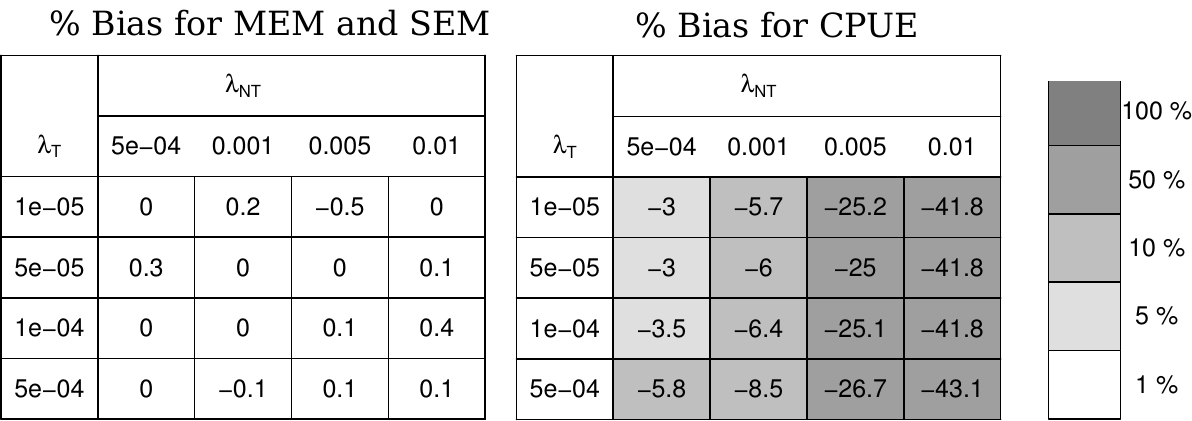}
\caption{Relative bias, defined as the absolute value of the bias divided by the true value $\vert\lambda_T-\hat{\lambda}_T\vert / \lambda_T$, expressed as a percentage averaged over 5000 simulations for 220 hooks per longline and 20 sets. Bias for SEM1, SEM2, MEM1 and MEM2 is the same in this case. On the right the bias computed for CPUE indices shows that the bias  increases with the increase of relative abundance of the non-target species and also with the relative abundance of target species.}
\label{Fig:2-BiasnoEmpty}%
\end{figure}
\end{center}

Figure \ref{Fig:2-BiasnoEmpty} presents the bias for the two indices in the absence of empty hooks. SEM, MEM1 and  MEM2 produce exactly the same results because there are no empty hooks and the analytical formula has been used.

The estimations are unbiased for abundance indices built with the exponential model and so competition is effectively taken into account. On the other hand bias in the CPUE index  increases with increasing relative density of the non-target species. This result confirms that the CPUE index strongly depends on competition and should be avoided. This behavior is always the same in all situations which have been addressed in this simulation study. 

The coefficient of variation presented in Table  \ref{Tab:CVnoEmpty} depends on the number of data points relying on $N$ and $L$ but also on the relative abundance, parameters $\lambda_T$ and $\lambda_{NT}$. This coefficient could be very high for low relative abundance situations.

\begin{figure}[htbp]
\includegraphics[width=9cm, height=7cm]{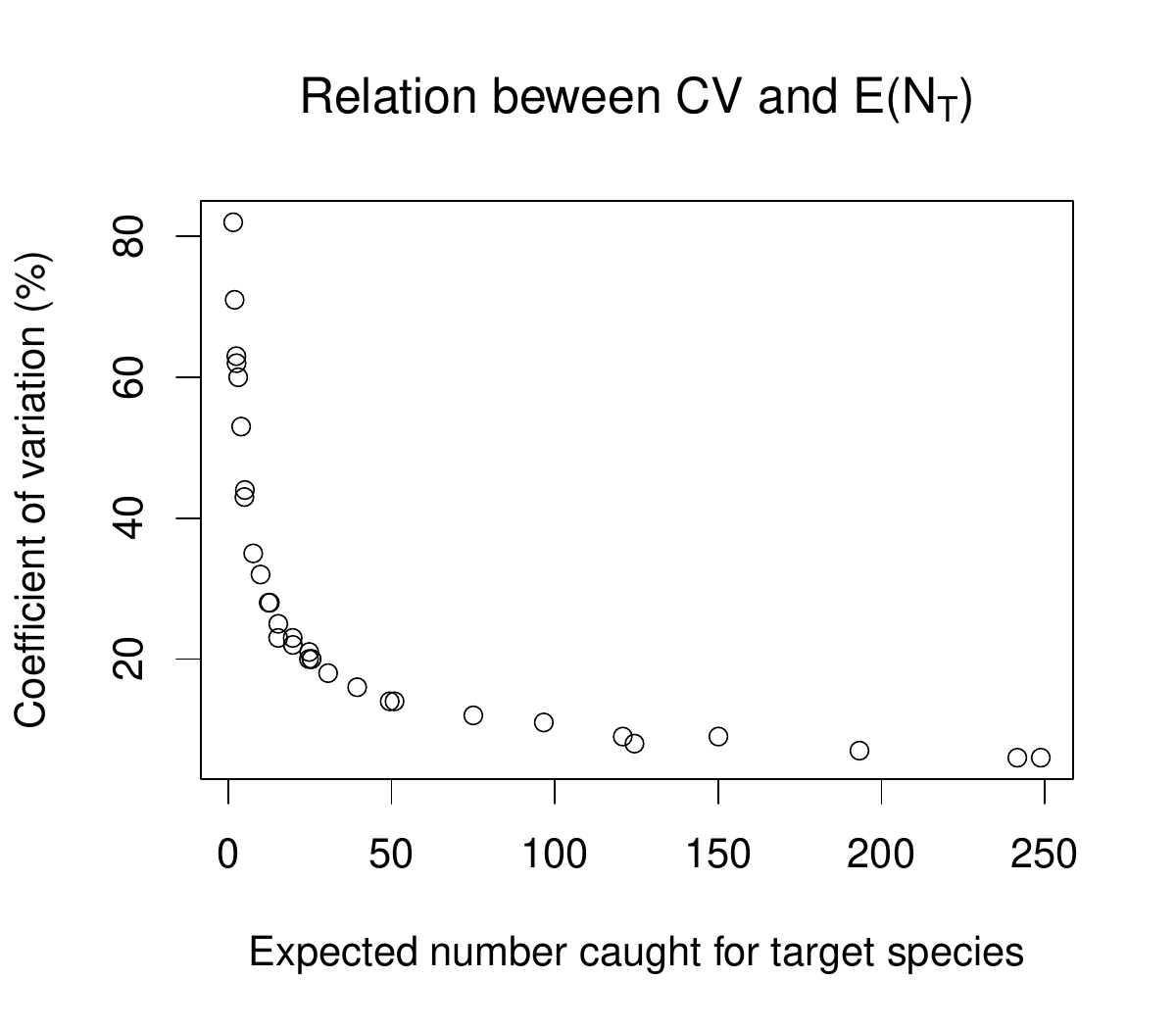}
\caption{Coefficient of variation for the estimates of $\lambda_T$ for MEM1 computed over 5000 simulations as a function of the expected number of catch for target species}
\label{Fig:3-CVnoEmpty}%
\end{figure}

\medskip

The coefficient of variation of the estimators for the relative abundance decreases with the expected number caught, which is in this study, $L\, N \frac{\lambda_T}{\lambda}\left( 1- e^{-\lambda S}\right)$. The relation between these two quantities is illustrated by Figure \ref{Fig:3-CVnoEmpty}. This expected number caught  describes the actual available information on $\lambda_T$.

The bias study shows that CPUE behaves poorly when interspecific competition occurs. In the following the results concerning this index will not be shown. 

\subsubsection{Competition and empty hooks}
SEM and MEM have the same behaviour when no empty hooks are present in the dataset. Simulations of empty hooks allow the comparison of  the respective behavior of those indices in presence of empty hooks. In our simulation context, the soak time and the initial number of hooks are constant in all sets, so that an  analytical formula can be used to compute the indices and SEM and MEM1 produce the same results.

\begin{center}
\begin{figure}[htbp]
\includegraphics[width=11cm]{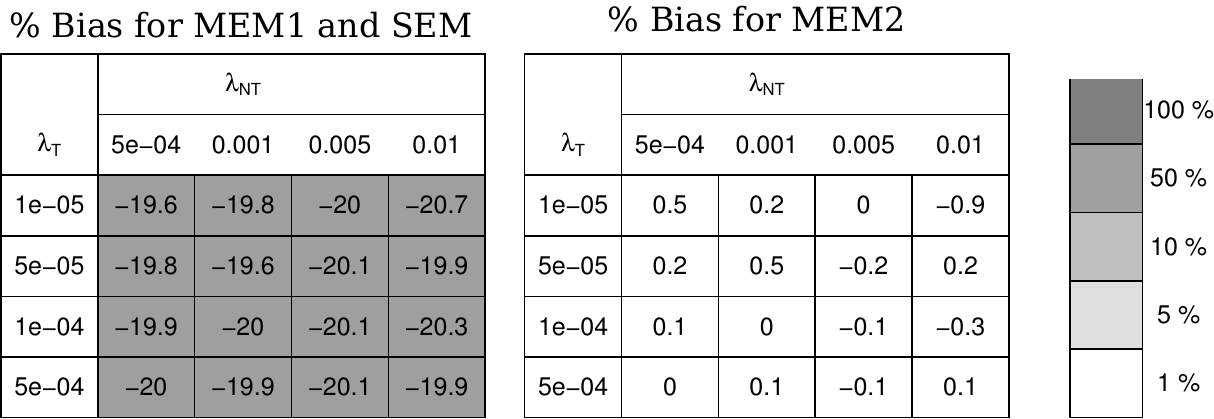}
\caption{Simulations under Scenario 2. MEM1, SEM1 and SEM2 produce the same results and tend to underestimate the relative abundance. MEM2 is the "true" model in this situation and tends to produce unbiased estimates.}
\label{Fig:4-BiasUniform}%
\end{figure}
\end{center}

Since MEM1 and MEM2 relies on two different hypotheses concerning the origin of the empty hooks, they give some different results. When all species are equally good at escaping (Sc.2) MEM2 produces unbiased estimators of the relative abundance while SEM1, SEM2 and MEM1 tend to underestimate this relative abundance. In the simulations used to produce Figure \ref{Fig:4-BiasUniform}, the probability of escape for non-target and target fish has been set to $20\%$, which corresponds to the underestimation of $20\%$ for the relative abundance $\lambda_{T}$.

\begin{center}
\begin{figure}[htbp]
\includegraphics[width=11cm]{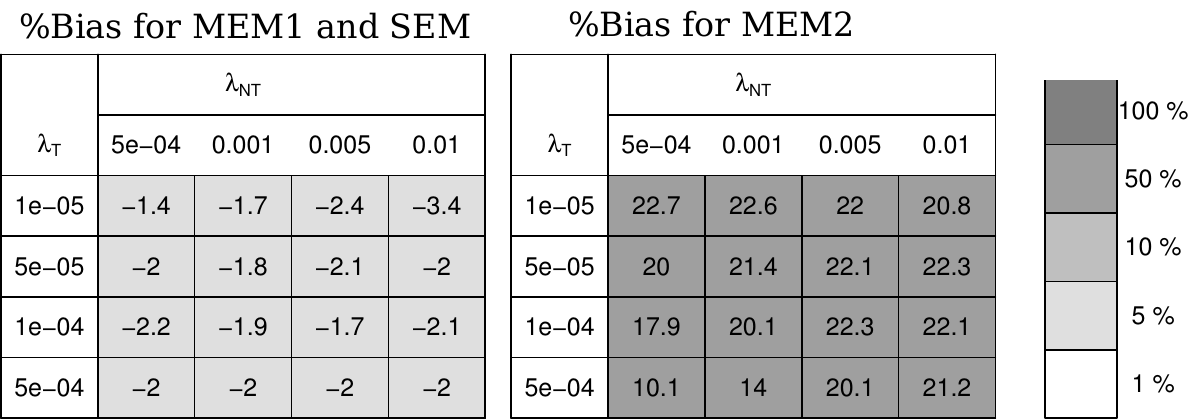}
\caption{Simulations under Sc1: MEM1 and SEM underestimate the true relative abundance by exactly the probability of escape for non-target species. MEM2 considerably overestimates the relative abundance.}
\label{Fig:5-Pref}%
\end{figure}
\end{center}

When the non-target species are better at escaping, MEM2 tends to overestimate the relative abundance of the target species. Results presented in Figure  \ref{Fig:5-Pref} are produced when $p_T=0.02$ and $p_{NT}=0.2$. MEM2 attributes a proportion of the empty hooks to the target species, this proportion depends on $\lambda_T$ and $\lambda_{NT}$. Since in this simulation most of the empty hooks arise from the non-target species, the higher the relative abundance of non-target species is,  the more the relative abundance of target species is overestimated. The bias in the estimates for MEM1 is constant and equals $2\%$ which corresponds to  the missed fish from the empty hooks.

\subsubsection{Numerical instability}
In order to study the numerical instability of the optimization algorithm, the maximum likelihood estimators have been computed through the analytical formula and using a numerical optimisation algorithm on the same data set (with shared $S$ and shared $N$).
Whichever scenario is used, the optimisation algorithm behaves well, i.e, there was less than $5\%$ of difference between the estimates computed using the analytical formula and the estimates obtained by numerical optimization except when the ratio between the relative abundance of non-target species $\lambda_{NT}$ and the relative abundance of the target species $\lambda_T$ was very high. In the extreme case, with $\lambda_{NT}=0.01$ and $\lambda_T=1e-05$ the average difference between the two estimates varies from  $10\%$ to $40\%$. This poor behaviour occurs when the log-likelihood peak is not strong enough. Some examples of this behaviour concerning the optimisation step are illustrated on the real data in section \ref{sec:RockfishResults}.

\subsection{Rockfish survey results}
\label{sec:RockfishResults}

\begin{center}
\begin{figure}[htbp]
  \includegraphics[width=17cm, height=8cm]{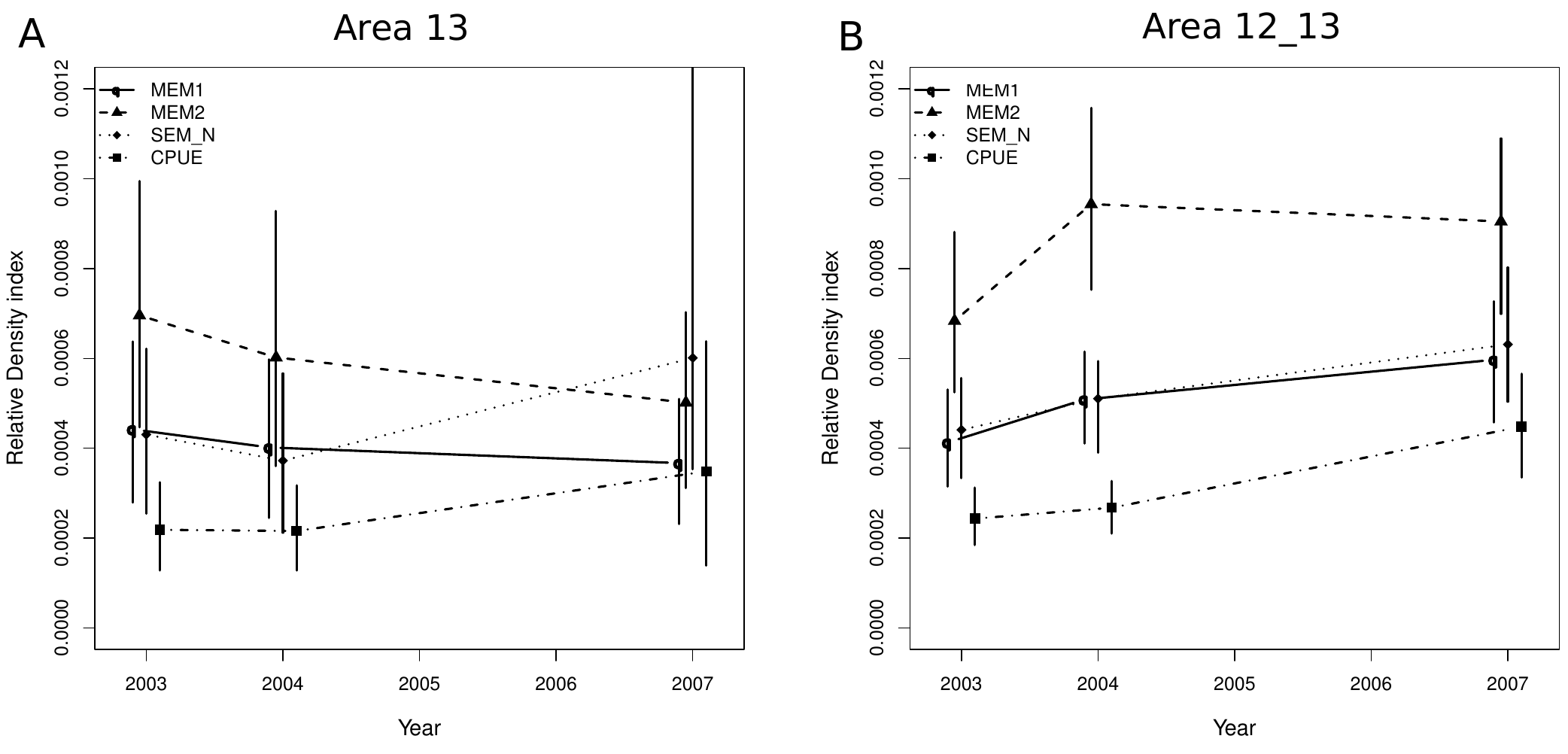}
\caption{Four indices computed for Area 13 (A on the left) and for Area 12 and 13 pooled (B on the right) for the quillback population. The numerical optimisation has some stability issues for year 2007, the results should be the same then MEM1 estimates. }
\label{Fig:6-Area13}
\end{figure}
\end{center}

Figure \ref{Fig:6-Area13}.A shows the different relative abundance indices obtained using the scientific survey described in section \ref{sec:presdata}. The confidence intervals have been computed using a bootstrap procedure with 5000 resamples.

The estimate for the numerical version of SEM index exhibits a considerable difference in trends and the confidence interval associated with this estimate is huge. This is due to a numerical instability problem. Indeed, there are very few quillback caught in 2007 in Area 13 as suggested by the decrease in the trend for all other indices. Therefore, the numerical optimization procedure behaves poorly which produces a poor estimate but also a large variability using a bootstrap procedure.
  
\begin{center}
\begin{figure}[htbp]
  \includegraphics[width=12cm, height=8cm]{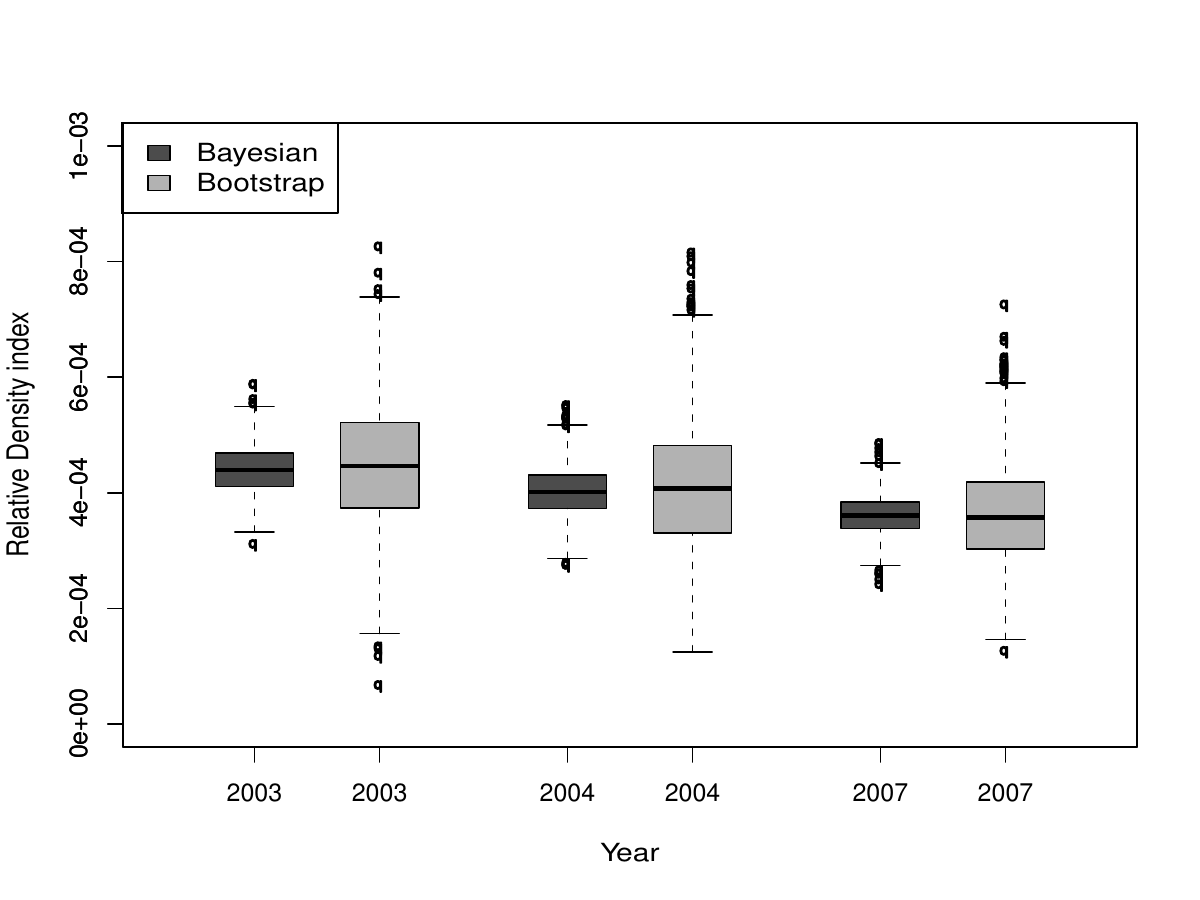}
\caption{Comparison between the variability given by bootstrap approach and the variability given by the posterior distribution in a Bayesian framework.}
\label{Fig:7-BoxPlot}
\end{figure}
\end{center}

The variability of the indices (see figure \ref{Fig:7-BoxPlot}) computed by a bootstrap approach is wider than the variability deduced from the posterior distribution. This point is discussed in the next section.

\begin{center}
\begin{figure}[htbp]
  \includegraphics[width=14cm, height=7cm]{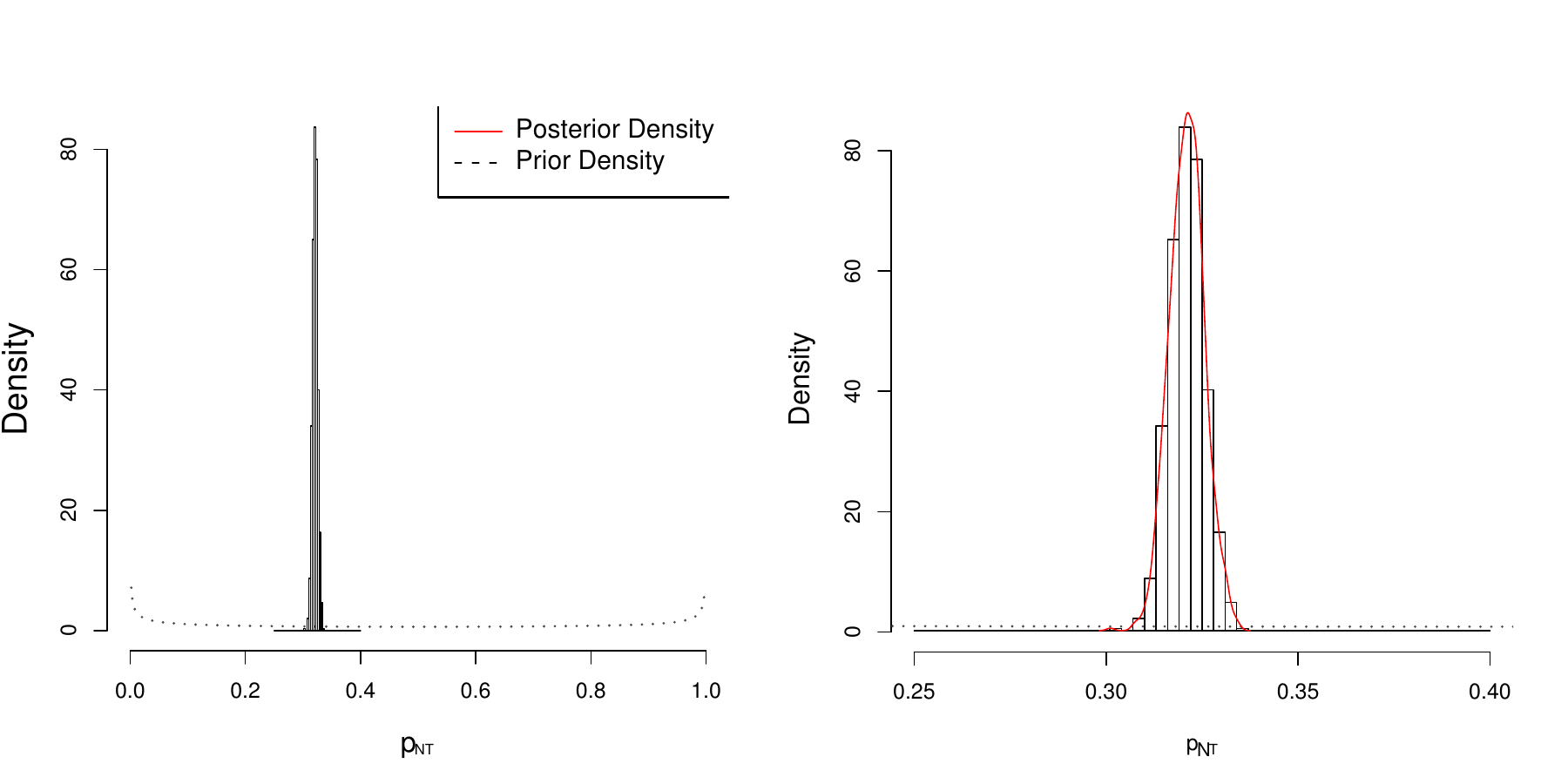}
\caption{Comparison between prior distribution and posterior distribution for parameter $p_{NT}$ in model MEM1 applied on the quillback population for the Strait of Georgia. The right panel is an elargement of the left one.}
\label{Fig:8-PosteriorPe}
\end{figure}
\end{center}

Another parameter of interest is the probability of escape, since this parameter measures the efficiency of the gear. Figure \ref{Fig:8-PosteriorPe} shows the posterior distribution of the parameter $p_{NT}$ estimated in MEM1. The posterior mean of $p_{NT}$ is $0.32$ with a standard deviation $4.4\, 10^{-3}$.
 
Within the Strait of Georgia survey, the northern Areas (12 and 13) have been surveyed together in the same years (2003, 2004 and 2007) and have alternated with surveys in the southern Areas (14-20). Since management isn't applied at the Area scale, data from Area 12 and 13 have been pooled to form one dataset. The corresponding relative abundance time series is shown in Figure \ref{Fig:6-Area13}.B.

Pooling the data from both Areas produces very similar trends in each of the indices and more precise estimates as shown in Table \ref{tab:CVquillback} and solves the numerical instability for the SEM index. The coefficient of variation is divided by almost two when using the whole dataset. We didn't include the other Areas of the Strait of Georgia (14-20) in the study since they were only surveyed once in 2005.

Given the uncertainty on the relative abundance indices, no change is statistically significant in the relative abundance of quillback population between years in the strait of Georgia. Even if the points estimates tend to show some increase we can't conclude that there has been any recovery of quillback stock because of the  uncertainty on the relative abundance estimates. The same study has been conducted for the yelloweye rockfish ({\it Sebastes ruberrimus}) showing that the overall trend tends to exhibit a small decrease but with no statistical significance.

\section{Discussion}
\label{sec:Discussion}
The first and most important conclusion of this study is that when interspecific competition was accounted for in simulated data, the classical longline CPUE estimators gave strongly biased estimates of stock trends in all cases evaluated. In contrast, the exponential model-based estimators showed much less bias. Therefore it is advisable to avoid the use of classical CPUE, at least from longline data, since this index doesn't take competition into account. From one year to another, the level of competition can vary and two CPUE indices computed during two different years are not comparable which is unacceptable for a relative abundance index. Even if the standard CPUE for the case studies are not so different from the relative abundance indices obtained with the alternative approaches, the simulations show that it is a major concern and that it could be very important for other longline experiments.

In the absence of empty hooks, the abundance indices built on SEM or MEM are  comparable even if the fundamental assumptions of SEM and MEM are very different. If $N_{T}$ is considered as a sum over all hooks of the number of hooks which have caught a target individual, the central limit theorem claims that $N_T$ exhibits a normal distribution provided that the number of hooks is large enough. The central limit theorem does not required the assumption on the independence of hooks, weak dependence, as Markovian dependance for example,  is sufficient (see \citet{Billingsley95} for a discussion of central limit theorem under weak dependence conditions). Therefore, the assumption of normality used in SEM is justified but the SEM assumes the independence beetween $N_T$ and $N_{NT}$. However, as the result of the competition, these two quantities are naturally anticorrelated and the asummption is obviously unrealistic. In contrast, MEM models  the dependence between $N_T$ and $N_{NT}$ but assumes the independence of hooks.

Ignoring empty hooks will produce poor indices because most of the empty hooks result from fish escapement which should  be accounted for in the relative abundance index. Empty hooks are of major importance to building abundance indices even if there is no perfect solution to deal with them. Biological knowledge can be very useful for deciding how to deal with empty hooks and the Bayesian framework offers an intuitive way to use this kind of information to remove the non-identifiability problem. But this biological knowledge is hard to obtain since the escapement of fish from hooks is difficult to study. 
MEM1 and MEM2 require fully explicit choices concerning the empty hooks. Even if this required choice is hard to make, the explicit choice should still be considered. In the SEM model the choice is made by default and it is even not explicit. We recommend that practitioners design studies to collect information about the ability of target and common non-target species to escape and work in a Bayesian framework using the code provided in Annex \ref{an:Winbugs2}.  The major advantage is that the uncertainty about empty hooks is translated to uncertainty about the relative abundance indices. If absolutely no information is available, the recommendation is to use SEM1 since the bias on the relative abundance index doesn't depend on the abundance of non-target species.
Furthermore, multiplicative constant bias as seen for MEM1 doesn't produce biased estimates of population dynamic parameters model since the multiplicative constant is absorbed by the coefficient of proportionality which links abundance indices and biomass.

One major drawback of all of the models formulated in this paper is the assumption of constant relative abundance along the longline set and independence between hooks which are obviously not true. This assumption of independence between the hooks may explain the lower uncertainty on the estimates of the relative abundance indices produced by the posterior distribution (figure \ref{Fig:8-PosteriorPe}). Assuming the independance between the hooks, we assume that the dataset is more informative than it really is. Furthermore \cite{Sigler00} has shown that  the catch rate decreases with time for sablefish ({\it Anoplopoma fimbria}) which tends to prove that the assumption of constant relative abundance is not always satisfied.  Different approaches could be explored to avoid this assumption. The first one would be to record the change of habitat along the longline set and using this as a covariate in the model. Another possibility would be to refine the modeling of the abundance index. For instance, we could consider a local relative abundance index $\lambda_{Th}$ at hook $h$ defined as the sum of the main relative abundance $\lambda_T$ plus a noise term. The noise term would be chosen as an autoregressive model for example to use the information of the hooks in the neighbourhood. This extra variability could account for the variability in the habitat. Another perspective for dealing with the variability along the longline is to define the abundance index as piecewise constant function along the longline and trying to detect the change in this function using the tools of change-point detection \citep{Lavielle01}.

The possible variation of $\lambda_T$ during the soak time is also a question of interest. Some species could be more attracted by a fresh bait and therefore $\lambda$ would be supposed to decrease with time. This question of attractivity of the baits has been studied by \cite{Ferno+95}  but currently there is no solution for taking this into account when building some abundance indices.

\section*{Acknowledgement}
The authors thank Andrew Edwards and Etienne Rivot for their useful advices and comments  for the writing of this paper. 
\bibliographystyle{apalike}
\bibliography{biblio}

%%%%%%%%%%%%%%%%%%%%%%%%%%%%%%%%%%%%%%%%%%%%%%%%%%%%%%%%%%%%%%%%%%%%%%%%%
%
% APPENDIX
%
%%%%%%%%%%%%%%%%%%%%%%%%%%%%%%%%%%%%%%%%%%%%%%%%%%%%%%%%%%%%%%%%%%%%%%%%%

\appendix
\section{CPUE computed from several observations}
\label{sec:cpue}
A CPUE index may be considered as the estimation of the probability of successes in a binomial trial. If $N_T$ stands for the number of success and $p$ is the probability of success over $N \times S$ trials, then
\begin{equation*}N_T\sim\mathcal{B}(p, N\times S).\end{equation*}
The maximum likelihood estimator $p$ is given by $N_T/(N\times S)$ =CPUE.  If  independent records $N_{T_l}$ for L longline sets are available with distribution $N_{T_l}\sim\mathcal{B}(p,N_l\times S_l)$, a sufficient statistic is the sum of all the catch $N_T^+$ which is distributed as
\begin{equation*}\sum_l N_{T_l} =N_T^+ \sim \mathcal{B} (p, \sum_{l} N_l \times S_l).\end{equation*}
The maximum likelihood estimator is defined as
\begin{equation*}\frac{\sum_{l=1}^L N_{T_l}}{\sum_{l}  S_l N_l},\end{equation*}
which defines a generalized CPUE definition.
\section{More details about the MEM}
\subsection{Likelihood of MEM with empty hooks}
\label{an:Distribution}

This section gives the key steps of the calculus for the likelihood given in formula \ref{eq:MEMlogLike}. We  use brackets to denote \emph{pdf}'s
as many conditioning terms will appear in the probabilistic
expressions derived from the full version of Multinomial Exponential Model. As in   \citet{gelfand+90},  the brackets denote either a density or
a discrete probability distribution.

The likelihood is then defined by:

\begin{align*}
l(\lambda_{T}, \lambda_{NT}, p_{T}, p_{NT})&=\left[ N_B, N_{T}, N_{NT}, N_{E}\vert  \lambda_{T}, \lambda_{NT}, p_{T}, p_{NT} \right],\\
&= \left[N_B\vert  \lambda_{T}, \lambda_{NT}\right]\, \left[N_{T},  N_{NT}, N_{E}\vert N_{B},   \lambda_{T}, \lambda_{NT}, p_{T}, p_{NT}\right].
\end{align*}

By definition of the model $\left[N_B\vert  \lambda_{T}, \lambda_{NT}\right]$ is a binomial distribution. We need then to define the joint distribution of $(N_T, N_{NT}, N_{E})$ given $N_B$ the total number of unbaited hooks. To obtain this distribution, we need to explicit the integration term over the hidden quatities $N_{T}^{(E)}$ and $N_{NT}^{(E)}$. To make  the writing easier to follow all the parameters will be omitted in the conditioning term.

\begin{align}
\nonumber \left [ N_T, N_{NT}, N_E \vert N_B\right]=& \sum_{k=0}^{N_E} \left[ N_T, N_{NT}, N_T^{(E)}=k, N_{NT}^{(E)}=N_E-k \vert N_B\right]\\
\nonumber=& \sum_{k=0}^{N_E} \Comb{N-N_B}{N_T+k} \left(\frac{\lambda_T}{\lambda}\right)^{N_T+k} \left(\frac{\lambda_{NT}}{\lambda}\right)^{N_{NT}+N_E-k} \Comb{N_{T}+k}{k} p_{T}^k (1-p_T)^{N_T} \\
\nonumber&\hspace{3cm} \Comb{N_{NT}+N_E - k}{N_E-k } p_{NT}^{N_E-k} (1-p_{NT})^{N_{NT}}\\
\nonumber=& \left\lbrace \left(\frac{\lambda_T}{\lambda}\right)\left(1-p_T\right)\right\rbrace^{N_T}\left\lbrace \left(\frac{\lambda_{NT}}{\lambda}\right)\left(1-p_{NT}\right)\right\rbrace^{N_{NT}}\frac{(N-N_B)!}{N_T!\, N_{NT}! }\\
\nonumber & \hspace{3cm}\sum_{k=0}^{N_E}\frac{1}{(N_E-k)!\, k} \left(\frac{\lambda_T}{\lambda} p_T\right)^{k} \left(\frac{\lambda_{NT}}{\lambda}p_{NT}\right)^{N_E-k} \\
\nonumber=& \frac{(N-N_B)!}{N_T ! \, N_{NT}!\, N_E ! }\left( \frac{\lambda_T}{\lambda}(1-p_{T}\right)^{N_T}\left( \frac{\lambda_{NT}}{\lambda}(1-p_{NT}\right)^{N_{NT}}\left( \frac{\lambda_{T}p_T + \lambda_{NT}p_{NT}}{\lambda}\right)^{N_{E}}
\end{align}

The likelihood is then obtained by combining the binomial distribution of $N_B$ with this previous result to give equation \ref{eq:MEMlogLike}.

\subsection{Identifiability}
\label{an:Identifiable}
To prove that the full version of the MEM is not identifiable it is sufficient to express the likelihood with only three parameters. Let us define $\alpha=\frac{\lambda_T}{\lambda}(1-p_T)$ and $\beta= \frac{\lambda_{NT}}{\lambda}(1-p_{NT})$. Therefore the likelihood given in equation \ref{eq:MEMlogLike} may be rewritten as:
\begin{equation}
l(\lambda,\alpha,\beta)=\frac{N!}{N_B !\, N_T ! \, N_{NT}!\, N_E ! } \left(e^{-\lambda S}\right)^{N_B}  \left(1-e^{-\lambda S}\right)^{N-N_B} \left( \alpha\right)^{N_T}\left( \beta\right)^{N_{NT}}\left( 1-\alpha-\beta \right)^{N_{E}}.
\label{eq:Regulier}
\end{equation}

This form of the model is identifiable and it is called the regular form of the model.

\subsection{Estimators}
\label{an:Estimations}
\subsubsection{MEM Estimators}
Using the regular form of the model given in equation \ref{eq:Regulier}, it is easy to derive the log likelihood and obtained the maximum likelihood estimators:
\begin{align*}
\hat{\lambda}=\frac{1}{S}\log{\left(\frac{N}{N_B}\right)},\quad
\hat{\alpha} = \frac{N_T}{N_T+N_{NT}+N_E},\quad
\hat{\beta} = \frac{N_{TN}}{N_T+N_{NT}+N_E}.
\end{align*}

Those estimators are quite intuitive: for example the $\alpha$ parameter represents the relative density of catchable individuals from target species and it is  estimated as the ratio between the the catch of target species and the total number of catch including empty hooks. The ambiguity of the model lies then in the definition of the relative density and how to link the relative density of catchable individuals to the actual relative density. In other words, how much the relative density of catchable individuals has to be increase to take into account the escaped individuals.

The maximum of the logllikelihood function equals
$$\log{\left(\frac{N!}{N_B! \, N_T !\  N_T! \ N_{NT}! \, N_{NE}}\right)}+\log{\left(\frac{N_B^{N_B} \, N_T^{N_T }\  N_T^{N_T} \ N_{NT}^{N_{NT}} \, N_{E}^{N_{E}}}{N^N}\right)} $$

This value could be used to compute an AIC criterion.
\bigskip

In the following we will derive the maximum likelihood estimators in our two specials cases of interest.

\subsubsection{MEM1 Estimators}
MEM1 corresponds to the assumption that the target species can't escape, so taht $p_T=0$, in this context the log likelihood is given by
\begin{align*}
L\left(\lambda_T,\lambda_{NT}, p_{NT}\right) = & K - N_B S \lambda + (N-N_B) \log{  \left(1-e^{-\lambda S}\right)} + N_T \log{(\lambda_T)}  \\
 &+ (N_{NT}+N_E) \log{(\lambda_{NT})}  + N_{NT} \log{(1- p_{NT})} \\
 & - (N-N_{B}) \log{(\lambda)} + N_{E}\log{(p_{NT})}
\end{align*}

The first derivatives are then derived:
\begin{align*}
\frac{\partial L}{\partial \lambda_T} = &  - N_B S + \frac{(N-N_B) S e^{-\lambda S}}{1-e^{-\lambda S}} + \frac{N_T}{\lambda_T} - \frac{N-N_B}{\lambda},\\
\frac{\partial L}{\partial \lambda_{NT}} = &  - N_B S + \frac{(N-N_B) S e^{-\lambda S}}{1-e^{-\lambda S}} + \frac{N_{NT} +N_E}{\lambda_{NT}} - \frac{N-N_B}{\lambda},\\
\frac{\partial L}{\partial p_{NT}} = &   - \frac{N_{NT}}{1-p_{NT}} + \frac{N_{E}}{p_{NT}}.\\
\end{align*}

The maximum likelihood estimators given on the left side of  equation \ref{eq:MLE} are then obtained by determining the roots of this set of equations.

If $N$ the number of hooks is large enough, the distribution of the estimators may be approximated by a normal distribution with mean $(\lambda_T,\lambda_{NT}, p_{NT})$ and a matrix variance which is the inverse of the Fisher information matrix $F$ defined by 
$$F=-\E \left[ \frac{\partial^2}{\partial \theta_1 \partial \theta_2} \right],$$
where $\theta$ represents a generic vector of parameters.

It can be shown that the asymptotic variance matrix for MEM1 is given by:
\begin{equation}
Cov_{MEM1} = \frac{1}{N}\left (
  \begin{array}{ccc}
    \frac{\lambda_T \lambda_{NT}}{1-e^{-\lambda S}} + \frac{1-e^{-\lambda S}}{S^2 e^{-\lambda S}} \frac{\lambda_T^2}{\lambda^2} &
    -\frac{\lambda_T \lambda_{NT}}{1-e^{-\lambda S}} + \frac{1-e^{-\lambda S}}{S^2 e^{-\lambda S}} \frac{\lambda_T\lambda_{NT}}{\lambda^2} & 0 \\
    -\frac{\lambda_T \lambda_{NT}}{1-e^{-\lambda S}} + \frac{1-e^{-\lambda S}}{S^2 e^{-\lambda S}} \frac{\lambda_T\lambda_{NT}}{\lambda^2} &    \frac{\lambda_T \lambda_{NT}}{1-e^{-\lambda S}} + \frac{1-e^{-\lambda S}}{S^2 e^{-\lambda S}} \frac{\lambda_{NT}^2}{\lambda^2} &  0 \\
    0 & 0 & \frac{\lambda p_{NT} (1-p_{NT})}{\lambda_{NT}(1-e^{-\lambda S})} \\
  \end{array}
\right)
\end{equation}

\subsubsection{MEM2 Estimators}
The equivalent equations for MEM2 (corresponding to the assumption that the probability of esacpe is the same for every species)with $p_{T}=p_{NT}=p$ can be obtained from the following LogLikelihood:
\begin{align*}
L\left(\lambda_T,\lambda_{NT}, p\right) = & K - N_B S \lambda + (N-N_B) \log{  \left(1-e^{-\lambda S}\right)} + N_T \log{(\lambda_T)}  \\
 &+ N_{NT} \log{(\lambda_{NT})}  + (N_{NT}+N_T) \log{(1- p)} \\
& - (N_{T}+N_{NT}) \log{(\lambda)} + N_{E}\log{(p)}
\end{align*}

The first derivatives are then derived:
\begin{align*}
\frac{\partial L}{\partial \lambda_T} = &  - N_B S + \frac{(N-N_B) S e^{-\lambda S}}{1-e^{-\lambda S}} + \frac{N_T}{\lambda_T} - \frac{N_T + N_{NT}}{\lambda},\\
\frac{\partial L}{\partial \lambda_{NT}} = &  - N_B S + \frac{(N-N_B) S e^{-\lambda S}}{1-e^{-\lambda S}} + \frac{ N_{NT} }{\lambda_{NT}} - \frac{N_{T}+N_{NT}}{\lambda},\\
\frac{\partial L}{\partial p} = &   - \frac{N_T + N_{NT}}{1-p} + \frac{N_{E}}{p}.\\
\end{align*}

The maximum likelihood estimators are obtained by determining the roots of this set of equations.

The asymptotic covariance matrix is then given by:
\begin{equation}
Cov_{\small MEM_2} = \frac{1}{N}\left (
  \begin{array}{ccc}
    \frac{\lambda_T \lambda_{NT}}{(1-p)(1-e^{-\lambda S})} + \frac{1-e^{-\lambda S}}{S^2 e^{-\lambda S}} \frac{\lambda_T^2}{\lambda^2} &
    -\frac{\lambda_T \lambda_{NT}}{(1-p)(1-e^{-\lambda S})} + \frac{1-e^{-\lambda S}}{S^2 e^{-\lambda S}} \frac{\lambda_T\lambda_{NT}}{\lambda^2} & 0 \\
    -\frac{\lambda_T \lambda_{NT}}{(1-p)(1-e^{-\lambda S})} + \frac{1-e^{-\lambda S}}{S^2 e^{-\lambda S}} \frac{\lambda_T\lambda_{NT}}{\lambda^2} &    \frac{\lambda_T \lambda_{NT}}{(1-p)(1-e^{-\lambda S})} + \frac{1-e^{-\lambda S}}{S^2 e^{-\lambda S}} \frac{\lambda_{NT}^2}{\lambda^2} &  0 \\
    0 & 0 & \frac{ p (1-p)}{1-e^{-\lambda S}} \\
  \end{array}
\right)
\end{equation}

\section{Some useful codes}
\label{an:Winbugs2}

\subsection{JAGS code for MEM1}
\label{sec:jagscodeMEM1}
\begin{verbatim}
var
# data
# Data is a matrix with 4 columns and NData lines
# Col 1 corresponds to Nb, Col2 to N1, Col3 to N2, col4 to Ne 
	NData, P[NData],  Data[NData,4], N[NData],
#variable requiring initialisation
		lambda1, lambda2,  p,
#variable without initialisation, deduced from the code
	alpha[NData, 4];

model {
/* prior density */
p ~ dbeta(0.1,0.1) ;
lambda1 ~ dbeta(0.1,0.1);
lambda2 ~ dbeta(0.1,0.1);
lambda  <- lambda1 + lambda2;
		
/* Model part */
for(j in 1:NData)
	{
		alpha[j,1] <- exp(-lambda * P[j]); 
		alpha[j,2] <- (1- exp(-lambda * P[j])) * lambda1 / lambda; 
		alpha[j,3] <- (1- exp(-lambda * P[j])) *lambda2/lambda *(1-p); 
		alpha[j,4] <- (1- exp(-lambda * P[j])) * lambda2 * p /lambda ; 
		Data[j,] ~ dmulti(alpha[j,], N[j]);
	}
}
\end{verbatim}

\subsection{JAGS code for MEM2}
\label{sec:jagscodeMEM2}
\begin{verbatim}
var
# data
# Data is a matrix with 4 columns and NData lines
# Col 1 corresponds to Nb, Col2 to N1, Col3 to N2, col4 to Ne 
	NData, P[NData],  Data[NData,4], N[NData],
# variable requiring initialisation
		lambda1, lambda2,  p,
# variable without initialisation, deduced from the code
	alpha[NData, 4];

model {
/* prior density */
p ~ dbeta(0.1,0.1) ;
lambda1 ~ dbeta(0.1,0.1);
lambda2 ~ dbeta(0.1,0.1);
lambda  <- lambda1 + lambda2;
		
/* Model part */
for(j in 1:NData)
	{
		alpha[j,1] <- exp(-lambda * P[j]); 
		alpha[j,2] <- (1- exp(-lambda * P[j])) * lambda1 * (1-p) / lambda; 
		alpha[j,3] <- (1- exp(-lambda * P[j])) *lambda2/lambda *(1-p); 
		alpha[j,4] <- (1- exp(-lambda * P[j])) * p ; 
		Data[j,] ~ dmulti(alpha[j,], N[j]);
	}
}
\end{verbatim}

\newpage

\begin{table}[htbp]
\begin{tabular}{|c|c|c|c|c|}\hline
  & \multicolumn{4}{c|}{ $\lambda_{NT} $}\\\cline{2-5} 
  $\lambda_T$ & 5e-04 & 0.001 & 0.005 & 0.01 \cr \hline
1e-05 & 43.2 & 44.8 & 50.9 & 56.7 \cr \hline
5e-05 & 19.5 & 20.1 & 22.2 & 25.3 \\ \hline
1e-04 & 13.8 & 14.4 & 15.9 & 17.9 \\ \hline
5e-04 & 6.2  &  6.4 &  7.4 &  8.1 \\ \hline
\end{tabular}
\caption{Coefficients of variation ($\%$) in estimates of $\lambda_T$ using the MEM models. Results For SEM, MEM1 and MEM2 are exactly the same since the soak time is shared by all longline sets and there are no empty hooks. The coefficients of variation have been computed over 5000 simulations for 220 hooks per longline and 20 sets.}\label{Tab:CVnoEmpty}
\end{table}

\clearpage 

\begin{table}
  \begin{tabular}{|c||c|c|c||c|c|c|}\hline
    \multicolumn{7}{|c|}{Coefficient of variation $(\%)$}\cr \hline 
      Index   & \multicolumn{3}{c||}{Area 13} & \multicolumn{3}{c|}{Area 12 and 13}\cr
      & 2003 & 2004 & 2007   & 2003 & 2004 & 2007 \cr  \hline
    MEM1 & $24.1$ & $26.6$ & $22.7$ &  $15.9$ & $12.6$ & $13.6$ \cr
    MEM2 & $23.5$ & $29$ & $23.1$ & $15.6$ & $13.6$ & $13.3$ \cr
    SEM N & $25.8$ & $30.2$ & $29.8$ & $15.6$ & $13.1$ & $13.9$ \cr
    CPUE & $26.4$ & $28.1$ & $24.7$ & $16$ & $13.4$ & $14.1$ \cr \hline
\end{tabular}
\caption{ Coefficient of variation of the different relative abundance indices expressed in percent. The variability strongly decreases when data are pooled together. The large value for the coefficient of variation for 2007 for the numerical version of SEM is due to optimization instability. }
\label{tab:CVquillback}
\end{table}

\end{document}